\documentclass[prl,reprint,showpacs,amsmath,amssymb,superscriptaddress,longbibliography]{revtex4-1}
\usepackage{graphicx,bm,color,mathptmx}
\usepackage[colorlinks=true,linkcolor=blue,citecolor=blue,urlcolor =blue]{hyperref}

\begin{document}

\title{Dispelling Rayleigh's Curse}

\author{J. Rehacek} 
\affiliation{Department of Optics, 
Palack\'y  University, 17. listopadu 12, 771 46 Olomouc, 
Czech Republic}

\author{M. Pa\'{u}r}
\affiliation{Department of Optics, 
Palack\'y  University, 17. listopadu 12, 771 46 Olomouc, 
Czech Republic}

\author{B. Stoklasa} 
\affiliation{Department of Optics, 
Palack\'y  University, 17. listopadu 12, 771 46 Olomouc, 
Czech Republic}

\author{L. Motka} 
\affiliation{Department of Optics, 
Palack\'y  University, 17. listopadu 12, 771 46 Olomouc, 
Czech Republic}

\author{Z. Hradil} 
\affiliation{Department of Optics, 
Palack\'y  University, 17. listopadu 12, 771 46 Olomouc, 
Czech Republic}

\author{L. L. S\'{a}nchez-Soto} 
\affiliation{Departamento de \'Optica, Facultad de F\'{\i}sica,
 Universidad Complutense, 28040~Madrid,  Spain} 
\affiliation{Max-Planck-Institut f\"ur die Physik des Lichts,
  G\"{u}nther-Scharowsky-Stra{\ss}e 1, Bau 24, 91058 Erlangen,
  Germany}

\begin{abstract}
  We devise a systematic method to determine the Fisher information
  required for resolving two incoherent point sources with a
  diffraction-limited linear imaging device. The resulting
  Cram\'er-Rao bound gives the lowest variance achievable for an
  unbiased estimator. When only intensity in the image plane is
  recorded, this bound diverges as the separation between the sources
  tends to zero, an effect which has been dubbed as Rayleigh's
  curse. However, this curse can be lifted using suitable coherent
  measurements.  In particular, we determine a class of measurements
  that can be easily implemented as projections on the orthogonalized
  derivatives of the point spread function of the system.
\end{abstract}


\maketitle

\emph{Introduction.---}
The spatial resolution of any imaging device is restricted by light
diffraction~\cite{Abbe:1873aa}, which causes a sharp point on the object
to blur into a finite-sized spot in the image. This information is
encoded in the point spread function (PSF)~\cite{Goodman:2004aa},
whose size determines the resolution: two points closer than the
PSF width will be difficult to resolve due to the substantial overlap
of their images. This is the physical significance of the famous
Rayleigh criterion~\cite{Rayleigh:1879ab}.

Needless to say, improving this limit is a source of continuing
research. Actually, in the past two decades a number of top-notch
techniques have appeared,  overcoming Rayleigh's limit under particular
conditions. They rely on nonconventional strategies, such as
near-field imaging or on nonclassical or nonlinear optical properties of
the object~\cite{Dekker:1997aa,Hell:2007aa,Kolobov:2007aa,
Natsupres:2009aa,Hell:2009aa,Patterson:2010aa,Cremer:2013aa}.  
However, these schemes are often challenging and require careful 
control of the source, which is not always possible.

Quite recently, Tsang and coworkers~\cite{Tsang:2015aa, Nair:2016aa,
  Ang:2016aa} have re-examined this question from the perspective of
estimation theory.  The idea is to use the Fisher information and the
associated Cram\'er-Rao lower bound (CRLB) to quantify how well the
separation between two poorly resolved incoherent point sources can be
estimated.  When only light intensity at the image plane is measured
(the basis of all the traditional techniques), the Fisher
information falls to zero as the separation between the sources
decreases and the CRLB diverges accordingly; this is Rayleigh's
curse~\cite{Tsang:2015aa}.  On the other hand, when the Fisher
information of the complete field is calculated, it remains constant and so
does the CRLB, which implies that the Rayleigh limit is subsidiary to
the problem.

These stunning predictions prompted a sequence of rapid-fire
experimental implementations~\cite{Sheng:2016aa,Yang:2016aa,
  Tham:2016aa,Paur:2016aa}. Nonetheless, these proposals have some
limitations, for the detection works only for small separations or is
limited to particular source profiles. Inspired by these developments,
we establish in this Letter a general procedure to construct optimal
measurements (see also the recent related work~\cite{Lupo:2016aa}). By
this, we mean spatial modes such that, when the signal is projected
onto them, they yield a constant Fisher information thus attaining the
CRLB; i.e., the separation can be estimated with the best achievable
precision.  

The key feature of these modes is the spatial symmetry.  Even more
interestingly, we determine detection schemes achieving the quantum
limit for any symmetric PSF: the associated modes turn out to be
orthogonal polynomials with respect to a measure that is just the PSF.

\emph{Model and measurements.---} 
We follow the basic model of Tsang and coworkers~\cite{Tsang:2015aa,
  Nair:2016aa, Ang:2016aa} and consider quasimonochromatic paraxial
waves with one specified polarization and one spatial dimension, $x$,
denoting the image-plane coordinate. We formulate what follows in a
quantum language, even though it can be directly applied to a
classical scenario.  A coherent complex amplitude $U( x )$ can be
assigned to a ket $| U \rangle $, such that
$U( x )=\langle x | U \rangle$, where $| {x} \rangle$ represents a
point-like source located at ${x}$.

We take a spatially-invariant imaging system.  The associated PSF,
which is just the normalized intensity response to a point light
source, is denoted as $I (x)=| \langle x | \Psi \rangle|^{2} = 
|\Psi (x)|^{2}$, where $\Psi (x)$ is the amplitude PSF, which we 
require to be inversion symmetric; i.e., $\Psi(x) = \Psi(- x)$, an 
assumption met by most aberration-free imaging systems.

Two incoherent point sources, each of the same intensity, are located
at two unknown points $X_{\pm} = \pm s/2$ in the object plane. This
regular configuration entails no essential loss of generality.  Our
objective is to estimate the separation
$\mathfrak{s} = X_{+} - X_{-}$.

The density matrix for the image-plane modes is thus
\begin{equation}
  \varrho_{\mathfrak{s}} =\tfrac{1}{2} ( 
  | \Psi_{+} \rangle \langle  \Psi_{+} |  + 
  | \Psi_{-} \rangle \langle  \Psi_{-} | ) \, ,
\end{equation}
where the spatially-shifted responses are
$ \Psi_{\pm} (x) = \langle x \pm~s/2 | \Psi \rangle$.  This density
matrix gives the normalized mean intensity profile:
$ \varrho_{\mathfrak{s}} (x) = \tfrac{1}{2} ( {|\Psi(x-s/2)|^2} +
|\Psi(x+s/2)|^2)$. The spatial modes excited by the two 
sources are not orthogonal, in general ($\langle \Psi_{-} | 
\Psi_{+} \rangle \neq 0$), which means that they cannot be 
separated by independent measurements. This is
the crux of the problem.

To estimate ${\mathfrak{s}}$ we must perform appropriate measurements.
Complete von Neumann tests~\cite{Peres:2002oz} will prove sufficient
for our purposes.  They consist of a set of orthonormal projectors
$\{ | {n} \rangle \langle {n} | \}$ (with
$ \langle {n} | {n^{\prime}} \rangle =\delta_{nn^{\prime}}$) resolving
the identity $\sum_n | {n} \rangle \langle {n} | = \openone$.  Each
projector represents a single output channel of the measuring
apparatus; the probability of detecting the $n$th output is given by
the Born rule
$p_n ({\mathfrak{s}}) =\langle {n} | \varrho_{\mathfrak{s}} |
{n}\rangle$.
The generalization to continuous observables is otherwise
straightforward.

The statistics of the quantum measurement carries information about
${\mathfrak{s}}$.  This is aptly encompassed by the Fisher
information~\cite{Fisher:1925aa,Petz:2011aa}, which is a mathematical
measure of the sensitivity of an observable quantity to changes in its
underlying parameters (the emitter's position).  It is defined as
\begin{equation}
  \label{eq:Fishdef1}
  \mathcal{F}_{\mathfrak{s}} = \mathbb{E} 
  \left [  \left ( 
      \frac{\partial \log p_{n} ({\mathfrak{s}})}{\partial {\mathfrak{s}}} 
    \right )^{2} \right ] \, , 
\end{equation}
with $\mathbb{E} [Y]$ being the expectation value of the random
variable $Y$. The Cram\'er-Rao lower bound
(CRLB)~\cite{Helstrom:1976ij,Holevo:2003fv} ensures that the variance
of any unbiased estimator $\hat{{\mathfrak{s}}}$ of the quantity
${\mathfrak{s}}$ is bounded by the reciprocal of the Fisher
information; viz,
\begin{equation}
  \label{eq:CRLB}
  \mathrm{Var} ( \widehat{{\mathfrak{s}}} \, ) \ge 
\frac{1}{\mathcal{F}_{{\mathfrak{s}}}} \, .
\end{equation}

Let us take the von Neumann measurement as the continuous projection
over $| x \rangle \langle x |$, which corresponds to conventional
image-plane intensity detection (or photon counting, in the quantum
regime). We stress that this is the information used in any
traditional technique, including previous superresolution approaches.
The Fisher information (per detection event) for this scheme reads as
\begin{equation}
  \label{eq:Fishcla}
  \mathcal{F}_{\mathfrak{s}}  = \int_{-\infty}^{\infty}
  \frac{1}{\varrho_{\mathfrak{s}} (x)} 
  \frac{\partial^{2} \varrho_{\mathfrak{s}}(x)}{\partial {\mathfrak{s}}^{2}}
  \, dx \simeq 
  {\mathfrak{s}}^{2} \int_{-\infty}^{\infty}
  \frac{[I^{\prime \prime} (x)]^{2}}{I (x)}   \, dx \, ,
\end{equation}
where, in the second integral we have performed a first-order
expansion in $s$, which is valid only for points sufficiently close
together. Then, $\mathcal{F}_{\mathfrak{s}} $ goes to zero
quadratically as ${\mathfrak{s}} \rightarrow 0$. This means that
detection of intensity at the image plane is progressively worse at
estimating the separation for closer sources, to the point that the
variance in this situation is doomed to blow up.

To bypass this obstruction, we need a different measurement
that incorporates the information available in the phase discarded by the
intensity detection. In what follows we require our measurement to
have a well-defined parity; i.e.,
$\langle - x | {n} \rangle = \pm \langle x | {n}
\rangle$. Accordingly,
\begin{equation}
\label{eq:echa}
  p_{n} \equiv |a_n|^2 =  | \langle {n} | {\Psi_{\pm}} \rangle |^2 \, ,
\end{equation}
so that the measurement does not feel the two-component structure of
the signal. Additionally, the probability amplitudes $a_{n}$  have to
fulfill 
\begin{equation}
  \label{eq:condition}
  \mathrm{Im} 
\left ( {a_n} \frac{\partial a_n^{\ast}}{\partial {\mathfrak{s}}} \right ) =0 \, .
\end{equation}
This property allows one to write the Fisher information as
\begin{equation}
  \label{eq:nexF}
  \mathcal{F}_{\mathfrak{s}} = 
  4    \sum_n \left|\frac{\partial a_n}{\partial {\mathfrak{s}}}  \right|^{2} \, .
\end{equation}

Next, we note that $| \Psi_{\pm} \rangle = \exp( \pm i s P/2) | \Psi
\rangle$. Here,  $P$ is the momentum operator that generates 
displacements in the $x$ variable, so it acts as a derivative 
$P = - i \partial_{x}$, much in the same way as in quantum optics. 
Because of the completeness of the eigenstates of $P$,
Eq.~\eqref{eq:echa} can be rewritten in the form 
\begin{equation}
  \label{an_def}
  a_n= \int \langle n | p \rangle \, 
\langle p \, | \Psi \rangle \, e^{-i  s p/2} dp \, .
\end{equation}
Inserting this in (\ref{eq:nexF}), performing the derivative, and   
using measurement completeness, we finally obtain the compact
expression 
\begin{equation}
  \label{final-fisher}
  \mathcal{F}_{\mathfrak{s}} = \int p^2 \, |\Psi(p)|^2  dp = 
 \langle P^{2} \rangle,
\end{equation}
where $\Psi (p) $ is the Fourier transform of the PSF amplitude $\Psi
(x)$. The Fisher information appears then as the second
moment of the momentum with respect to the PSF and is therefore
independent of the separation of the points. Consequently,  the
variance in the  CRLB remains constant and one lifts Rayleigh's curse,
as heralded. Incidentally, Eq.~\eqref{final-fisher}  is known to be
the  ultimate quantum limit~\cite{Braunstein:1994aa,Paris:2009aa}. 
Hence,  we have clearly identified conditions enabling a 
measurement to attain the quantum CRLB. This is the first main result
of this Letter. 

%%%%%%%%%%%%%%%%%%%%%%%%%%%%%%%%%%
\begin{figure}[t]
  \includegraphics[width=0.95\columnwidth]{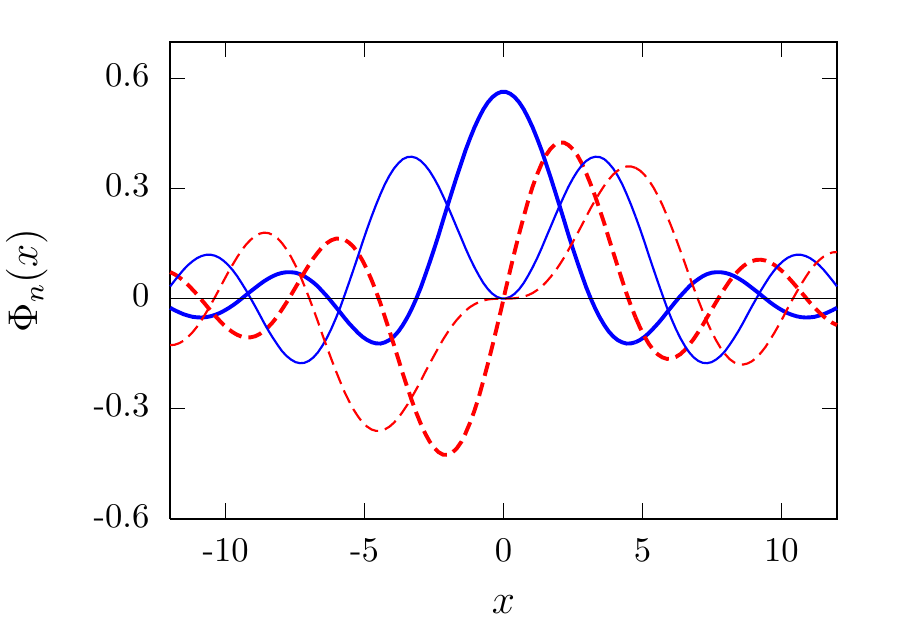}
  \caption{First PSF-adapted modes (\ref{sincderiv}) for a sinc
    response. In blue solid lines we plot the symmetric modes [$n=0$
    (thick) and $n=2$ (thin)] and in red broken lines we have the
    antisymmetric ones [$n=1$ (thick) and $n=3$ (thin)].}
  \label{fig:modes}
\end{figure}
%%%%%%%%%%%%%%%%%%%%%%%%%%%%%%%%%%
%%%%%%%%%%%%%%%%%%%%%%%%%%%%%%%%%%
\begin{figure*}[t]
  \includegraphics[width=0.68\columnwidth]{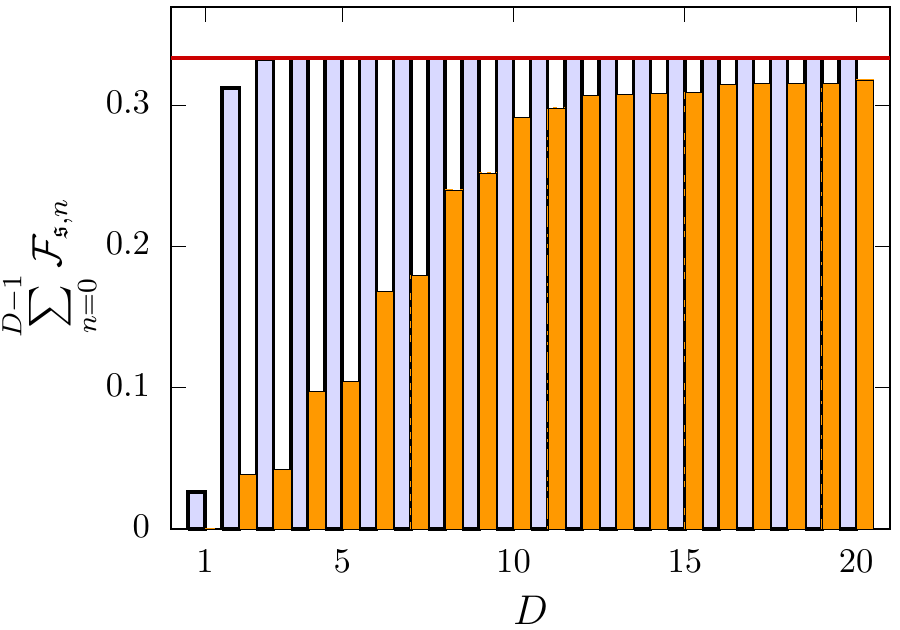}
  \includegraphics[width=0.68\columnwidth]{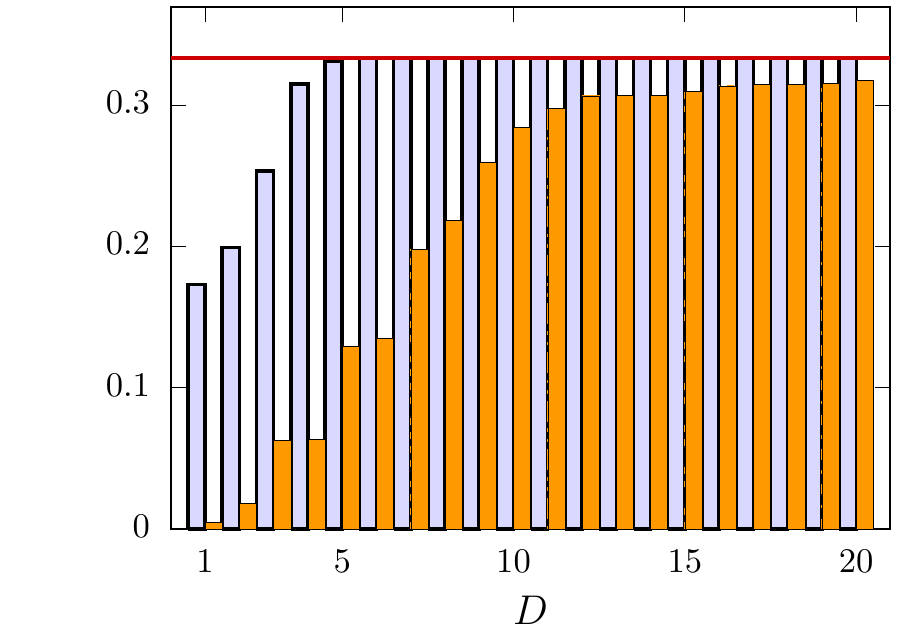}
  \includegraphics[width=0.68\columnwidth]{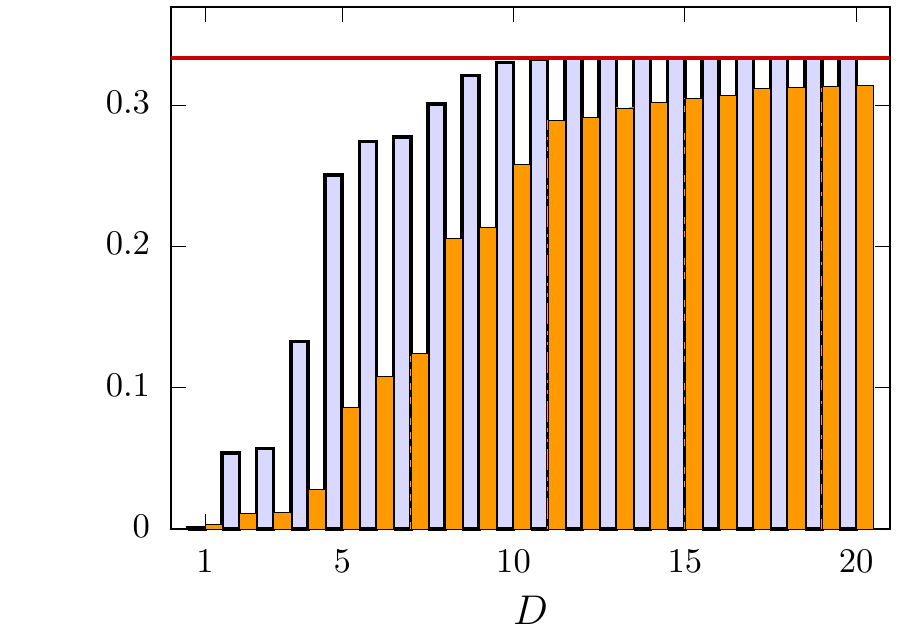}
  \caption{Fisher information attained by the first $D$ projections on
    the Hermite-Gauss basis with arbitrarily chosen $\sigma=\pi$
    (orange bars) and the PSF-adapted measurement \eqref{sincderiv},
    when applied to a system with a sinc impulse response.  The plots
    are for separations ${\mathfrak{s}} =1$ (left), ${\mathfrak{s}}=2$
    (middle), and ${\mathfrak{s}} = 15$ (right), and the corresponding
    Rayleigh limit is ${\mathfrak{s}}=\pi$.  More than a hundred of
    Hermite-Gauss projections must be measured to access $98.5\%$ of
    the quantum Fisher information (indicated by a horizontal red
    line) for ${\mathfrak{s}}=[0, 15]$, whereas just ten projections
    of the PSF-adapted measurement are sufficient.}
  \label{figcumul}
\end{figure*}
%%%%%%%%%%%%%%%%%%%%%%%%%%%%%%%%%%

Before proceeding further, we return to Eq.~\eqref{eq:condition}.
This condition is readily satisfied by choosing
$a_n= |{a}_n| \exp( i \alpha_n )$, with phases $\alpha_n$ independent
of ${\mathfrak{s}}$. As these phases can be absorbed in the basis
$|n\rangle$, this is tantamount to requiring real probability
amplitudes $a_n$. Due to the properties of the Fourier transform, this
imposes
\begin{equation}
  \label{eq:cond2}  
  \langle n | p \rangle \, 
  \langle p \, | \Psi \rangle = \pm \left(\langle n | p \rangle \, 
    \langle p \, | \Psi \rangle \right)^* \, ,
\end{equation}
where the symmetries of the PSF amplitude and the measurement have
been employed.  Hence a sufficient condition to enabling
\eqref{eq:condition} is that the diagonal elements of the operators
$ | \Psi \rangle \langle n |$ in the momentum representation are all
real or purely imaginary.  For a real PSF amplitude, an obvious choice of
$| n \rangle$ is a complete set of real wave functions $\langle x |
{n} \rangle$ with a well-defined parity (symmetric or antisymmetric).
The position projection (i.e., intensity detection) does not have the
required symmetry and this explains why the intensity scan fails to
reach the quantum bound.

\emph{Optimal strategies.---}
It is known that the optimal measurement in the limit of small
${\mathfrak{s}}$ is proportional to the first derivative of
$\Psi(x)$~\cite{Paur:2016aa}.  This suggests that one could try to
project the signal on a set of orthonormalized derivatives of
$\Psi(x)$.  Indeed, we propose to construct the measurement basis
$| {n} \rangle$ in momentum space as
\begin{equation}
\label{eq:lc}
\Phi_{n} ( p ) \equiv  \langle {p} | {n} \rangle = Q_n(p) \Psi(p),
\end{equation}
where $Q_n(p)$ is a system of orthogonal polynomials, with
respect to the measure $| \Psi (p) |^{2} dp$.  Since this measure is
symmetric, they  satisfy the symmetry property $Q_n(- p) = (-1)^{n}
Q_n( p)$~\cite{Szego:1938aa}.  

One can check that this generates a \emph{bona fide} measurement
basis.  Truly, for the states (\ref{eq:lc}), the condition
\eqref{eq:cond2} trivially holds, the probability amplitudes
$a_{n} = \langle n | \Psi_{\pm} \rangle$ are real, and 
\eqref{eq:condition} is fulfilled. Of course, one would expect that
the number of significant projections is small, and even the first
derivative is sufficient in the superresolution regime.

The optimal PSF-adapted modes attaining the CRLB (\ref{final-fisher})
for all separations are obtained by an inverse Fourier transform
\begin{equation}
  \label{eq:modes}
  \Phi_{n} (x) \equiv \langle x | n \rangle = 
  \frac{1}{\sqrt{2\pi}} \int Q_{n} (p) \Psi (p) e^{i p x } \, dp \, .
\end{equation}
The general rules \eqref{eq:lc} and \eqref{eq:modes} of finding the
PSF-optimized scheme make the second main result of this Letter.

As a first, important example, we consider a Gaussian PSF amplitude
$ \Psi(x) = (2\pi)^{-1/4} \exp(-x^2/4)$, with unit variance ($\sigma =
1$). The Fourier transform is again a Gaussian, and a direct calculation
gives $\mathcal{F}_{\mathfrak{s}} =1/4$. The optimal PSF-adapted set consists of
Hermite-Gauss polynomials, which are orthonormal with
respect to the PSF.

As a second example, we take a slit aperture with 
$\Psi(x)= \frac{1}{\sqrt{\pi}} \mathrm{sinc} (x)$ and Fourier transform
$\Psi (p) = \frac{1}{\sqrt{2}}  \mathrm{rect}(p/2)$. Here,
$\mathrm{sinc} (x)= \sin (x) / x$ and $\mathrm{rect}(p)$  is 0 outside
the interval $[-1/2, 1/2]$ and 1 inside it. The set $\Phi_{n} ( p )$ is
now the Legendre polynomials $L_{n} (p)$, which are complete in the unit
interval.  In this way,
\begin{equation}
  a_{n} = \langle n | \Psi_{\pm} \rangle = 
  \frac{\sqrt{2n+1}}{2}\int_{-1}^{1} L_n(p) \,  e^{-i  p s/2} \,  d p \, .
\end{equation}
By Eq.~\eqref{final-fisher}, the Fisher information is
$ \mathcal{F}_{\mathfrak{s}} = 1/3$ and, by \eqref{eq:modes}, the
optimal measurement modes are given as
\begin{equation}
  \label{sincderiv}
  \Phi_n (x)= \sqrt{n+1/2} \, \, \frac{J_{n+\frac{1}{2}}(x)}{\sqrt{x}},
\end{equation}
where $J_k(x)$ is the Bessel function of the first kind. For $n=1$,
the measurement reduces to the first derivative of the sinc function, as
expected. In Fig.~\ref{fig:modes} we plot the first PSF-adapted
modes~\eqref{sincderiv}. 

Each projection contributes with a piece of  information
\begin{equation}
  \mathcal{F}_{{\mathfrak{s}},n}=\frac{\pi \left [ 
  n J_{n-  \frac{1}{2}} \left ( {\mathfrak{s}}/2 \right ) -
   (n+1)  J_{n+\frac{3}{2}} \left ( {\mathfrak{s}}/2 \right ) \right ]^2}
 {(2 n+1) {\mathfrak{s}}} \, ,
\end{equation}
and, interestingly enough, all these complicated terms sum up to 
a total of $\mathcal{F}_{\mathfrak{s}}=1/3$. 

If the PSF amplitude of the system is real, any complete set of
real modes with defined parity will also work.  For example, the
sinc response will also be optimal with a projection on Hermite-Gauss modes,
but the connection to derivatives is lost. Besides, it might happen
that, for small separations, the number of significant projections 
will be much larger than for the PSF-adapted set.

This can be illustrated by comparing the performances of different
sets of measurements for the same PSF. Assuming a sinc, we can compare
two optimal sets: the PSF-adapted measurement of
Eq.~\eqref{sincderiv} and the Hermite-Gauss projections.
Figure~\ref{figcumul} shows the information obtained by summing the
Fisher information over the first $D$ projections.  Both measurements
attain the quantum CRLB; however, the number of effective projections
to be measured is considerably less for the optimized measurement. We
can also see that this advantage decreases with increasing
separations.

%%%%%%%%%%%%%%%%%%%%%%%%%%%%%%%%%%
\begin{figure}[t]
  \includegraphics[width=0.9\columnwidth]{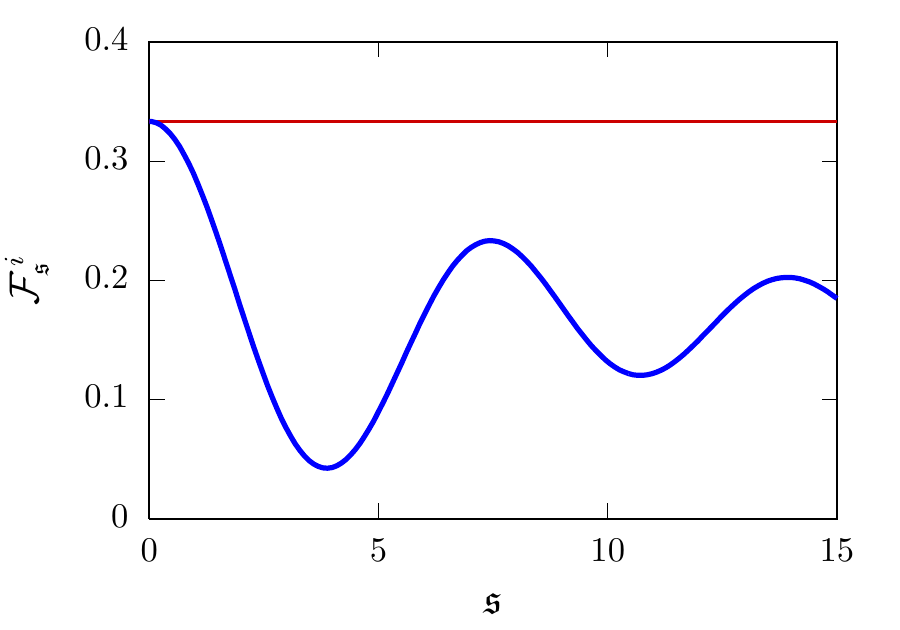}
  \caption{Fisher information accessed by measuring the Fourier sine
    transform of two incoherent images separated by a distance
    ${\mathfrak{s}}$ for a sinc impulse response.  Notice that for
    small separations this measurement is optimal. For larger
    separations the complement of
    $\mathcal{F}_{\mathfrak{s}}^{\mathrm{i}}$ to quantum Fisher
    information is accessed by the Fourier cosine transform.}
  \label{figimag}
\end{figure}
%%%%%%%%%%%%%%%%%%%%%%%%%%%%%%%%%%

Another feasible option is to project in modes comprised of the real
and imaginary parts of plane waves; i.e.,
\begin{equation}
  \Phi_{k}^{\mathrm{r}} (x) = \frac{1}{\sqrt{2\pi}} \cos(k x) \, ,
  \qquad
  \Phi_{k}^{\mathrm{i}} (x) = \frac{1}{\sqrt{2\pi}} \sin (k x) \, ,
\end{equation}
where now the modes are indexed by the continuous wave vector $k$.
This can be implemented using the Fourier properties of optical
lenses and spatial light modulators.  In Fig.~\ref{figimag}
we plot the Fisher information corresponding to 
\begin{equation}
    \mathcal{F}^{\mathrm{i}}_{\mathfrak{s}} = \tfrac{1}{2}
 \! \int\limits_{-1}^{1}
    k^{2} \sin^{2}\left ( \frac{k {\mathfrak{s}}}{2} \right ) \, d k \, , 
\end{equation}
obtained by such a measurement for different separations. For small
separations, the imaginary part of the signal spectrum alone readily
provides all the available information. Again, we note that in this
limit the same information can be extracted with a single projection
of the PSF-optimized measurement~\eqref{sincderiv}.

\emph{Concluding remarks.---}
In conclusion, we have shown that optimal sub-Rayleigh two-point
resolution can be achieved with an optical system having a symmetric
amplitude PSF provided the system output is projected onto a suitable
complete set of modes with definite parity.  Particularly useful modes
can be generated from the derivatives of the system PSF, which in the
limit of small separation can access all available information with a
single projection.

The above formalism can be generalized to other transformations
provided the frequency spectrum is replaced with a suitable
representation, in which the assumed transformation becomes a simple
phase shift.

\emph{Acknowledgments.---} 
We thank Gerd Leuchs, Olivia Di Matteo, and Matthew Foreman for
valuable discussions and comments. Encouraging exchanges with Mankei
Tsang are also appreciated.  We acknowledge financial support from the
Technology Agency of the Czech Republic (Grant TE01020229), the Grant
Agency of the Czech Republic (Grant No. 15-03194S), the IGA Project of
the Palack{\'y} University (Grant No. IGA PrF 2016-005) and the
Spanish MINECO (Grant FIS2015-67963-P).

%\bibliography{Resolution}

\end{document}